Computationally image-corrected dual-comb microscopy with a free-running single-cavity dual-comb fiber laser

Takahiko Mizuno[1,2,†], Yoshiaki Nakajima[2-4,†], Yuya Hata[2,3], Takuya Tsuda[2,5], Akifumi Asahara[2,3], Takashi Kato[2,3], Takeo Minamikawa, Takeshi Yasui[1,2,*], and Kaoru Minoshima[2,3,*]

[1]Institute of Post-LED Photonics (pLED), Tokushima University, 2-1 Minami-Josanjima, Tokushima, Tokushima 770-8506, Japan

[2]JST, ERATO, MINOSHIMA Intelligent Optical Synthesizer Project, 2-1 Minami-Josanjima, Tokushima, Tokushima 770-8506, Japan

[3]Graduate School of Informatics and Engineering, The University of Electro-Communications, 1-5-1 Chofugaoka, Chofu, Tokyo 182-8585, Japan

[4]Graduate School of Science, Toho University, 2-2-1 Miyama, Funabashi, Chiba 274-8510, Japan

[5]Graduate School of Advanced Technology and Science, Tokushima University, 2-1 Minami-Josanjima, Tokushima, Tokushima 770-8506, Japan

[†]These authors contributed equally to this article.

[*]Corresponding authors.

**Abstract (100 words)**

Dual-comb microscopy (DCM), an interesting imaging modality based on the optical-frequency-comb (OFC) mode and image pixel one-to-one correspondence, benefits from scan-less full-field imaging and simultaneous confocal amplitude and phase imaging. However, the two fully frequency-stabilized OFC sources requirement hampers DCM practicality due to the complexity and costs. Here, a bidirectional single-cavity dual-comb fiber laser (SCDCFL) is adopted as a DCM low-complexity OFC source. Computational image correction reduces the image blur caused by the SCDCFL residual timing jitter. Nanometer-order step surface profilometry with a 14.0 nm uncertainty highlights the image-corrected DCM effectiveness. The proposed method enhances the DCM versality and practicality.

# 1. Introduction

The ability to act as an optical carrier of amplitude and phase with a vast number of discrete, regularly spaced frequency channels is an interesting aspect of an optical frequency comb (OFC). While this aspect has been effectively applied for optical frequency rulers in optical frequency metrology and spectroscopy with the help of laser control [1-3], the combination of this aspect with space-to-wavelength conversion opens a new door to imaging applications for OFCs, namely, dual-comb imaging (DCI) [4-11]. In DCI, the image pixels to be measured are spectrally encoded into OFC modes by space-to-wavelength conversion or spectral encoding (SE) [13-16]. Then, the entire image is decoded at the same time from the mode-resolved spectrum of the image-encoded OFC acquired by dual-comb spectroscopy (DCS) [17-20] based on the one-to-one correspondence between image pixels and OFC modes. Due to the scan-less imaging capability in SE and the capability for simultaneous acquisition of amplitude and phase spectra in DCS, the combination of DCI with confocal laser microscopy, namely, dual-comb microscopy (DCM) [4, 5, 8, 11], enables scan-less confocal one-dimensional (1D) or two-dimensional (2D) imaging of amplitude and/or phase. For example, DCM has been effectively applied for surface topography of a nanometer-scale step-structured sample and nonstaining imaging of standing culture fixed cells [5]. Furthermore, DCM has been further expanded to confocal fluorescence microscopy featuring the scan-less fluorescence lifetime imaging [12], which is an important imaging modality in life sciences. However, the need for two fully frequency-stabilized OFC sources hampers the practical use of DCI and DCM due to their costs and complexity.

Recently, low-complexity OFC sources have been developed for versatile DCS. A quantum cascade laser (QCL) [21] is a chip-scale, high-power OFC source. While a pair of QCLs has been applied for DCS [22], this approach often suffers from poor mutual coherence between them. The microresonator soliton comb (microcomb) [23] is another chip-scale OFC source with better mutual coherence between the microcombs. A pair of microcombs has been used for DCS [24] and even DCI [9]. However, the relatively large repetition rate $f_{rep}$ corresponding to the frequency spacing significantly reduces the number of sampling points in the spectrum or the image. For example, the total number of image pixels was only several hundred in 2D images [9].

The ideal OFC source for versatile DCI and DCM has high mutual coherence to suppress image blur and moderate $f_{rep}$ to enable a sufficient number of 2D image pixels without the need for any frequency stabilization. One promising OFC source is a single-cavity dual-comb fiber laser (SCDCFL) [25-33]. In an SCDCFL, a pair of OFCs with slightly different repetition frequencies ($f_{rep1}$, $f_{rep2} = f_{rep1} + \Delta f_{rep}$) is generated from a single fiber cavity by multiplexing mode-locking oscillation in wavelength [25-27], polarization [28-30], or propagation direction [31-33]. Since the dual OFCs propagate through the same cavity, they experience almost the same cavity disturbances, and the resulting common-mode fluctuations prevent the decline in the mutual coherence between them under no active frequency stabilization. Furthermore, an $f_{rep}$ of approximately 100 MHz leads to tens-to-hundreds of thousands of OFC modes within the range of the optical spectrum, which is sufficient for the number of 2D image pixels. Although these SCDCFLs have been extensively applied for DCS, no attempts to apply them for DCI and DCM have been made.

In this article, we adopted a bidirectional SCDCFL [33] for versatile DCM. This bidirectional SCDCFL benefits from good spectral overlap over a wide spectral range, high stability and wide tunability of $\Delta f_{rep}$, and passive cancelation of common-mode noise. The image blur resulting from the residual timing jitter between the dual OFCs in the SCDCFL was computationally corrected by use of a self-reference image or external-reference image.

## 2. Experimental setup

Figure 1 shows a schematic drawing of the experimental setup for DCM. A bidirectional SCDCFL was used as a DCM light source. As the details of the bidirectional SCDCFL are given elsewhere [33], we briefly describe the laser here. Two independent mode-locking oscillations were achieved in clockwise-circulating light and counterclockwise-circulating light in a fiber ring cavity by nonlinear polarization rotation and two saturable absorber mirrors. Part of the noncommon optical path in the cavity suppresses the competition of the two mode-locking oscillations and enables independent tunability of $f_{rep1}$, $f_{rep2}$, and $\Delta f_{rep}$. The temperature of the fiber cavity was actively controlled by a combination of a thermistor and a Peltier heater. After optical amplification with a pair of erbium-doped fiber amplifiers (EDFAs), two counterpropagating output light beams from the SCDCFL, namely, CCW-OFC (center wavelength = 1550 nm, mean power = 190 mW, $f_{rep\_CCW}$ = 43,037,370 Hz) and CW-OFC (center wavelength = 1550 nm, mean power = 7 mW, $f_{rep\_CW}$ = 43,038,493 Hz, $\Delta f_{rep}$ = $f_{rep\_CCW}$ - $f_{rep\_CW}$ = 1,123 Hz), were used as a signal OFC and a local OFC, respectively.

The CCW-OFC beam passing through a beam splitter (BS) was fed into a two-dimensional SE (2D-SE) optical system [5, 14-16] including a virtually imaged phased array [34] (VIPA, free spectral range = 15.1 GHz, finesse = 110) and a diffraction grating (G, groove density = 1200 grooves/mm, efficiency = 90 %). Then, it was irradiated as a 2D spectrograph of CCW-OFC modes on a sample after passing through a pair of lenses (L1 and L2, focal length = 150 mm) and a dry-type objective lens (OL, NA = 0.25, working distance = 5.5 mm, focal length = 16 mm). A 1951 USAF resolution test chart with a positive pattern (Edmond Optics, Barrington, NJ, USA, #38-257, spatial frequency: 1.00 lp/mm ~ 228 lp/mm) was used as a sample. The reflection, absorption, scattering, and/or phase change of the CCW-OFC beam in the sample encode the image contrast into the amplitude and phase spectra of the 2D spectrograph. As the CCW-OFC beam from the sample passed through the same optical system in the opposite direction, each wavelength component of the spectrograph spatially overlapped as the signal-image-encoded CCW-OFC.

For computational image correction in DCM, we inserted a reference arm after the 2D-SE optics. A portion (= 50 %) of the CCW-OFC beam passing through L2 was separated by a combination of a half-wave plate (HWP) and a polarization beam splitter (PBS). Then, the separated CCW-OFC beam was irradiated as a 2D spectrograph of CCW-OFC modes on another test chart (negative type, Edmond Optics, Barrington, NJ, USA, #38-256, spatial frequency: 1.00 lp/mm ~ 228 lp/mm) after passing through a lens (L3, focal length = 200 mm) to encode the reference image of the separated CCW-OFC, namely, the reference-image-encoded CCW-OFC. The reference-image-encoded CCW-OFC beam was combined with the signal-image-encoded CCW-OFC beam with a time separation of 2.67 ns by the PBS, and both were fed into the DCS experimental setup.

The optical bandwidth of CW-OFC for use as the local OFC in DCS was reduced by an optical bandpass filter (OBPF, center wavelength = 1556 nm, transmission passband = 6 nm) to avoid the aliasing effect in DCS. Then, the CW-OFC beam was spatially overlapped with the combined signal-image-encoded and reference-image-encoded CCW-OFC beam in a 90:10 single-mode fiber coupler (FC). The optical power ratio of the CCW-OFC beam to the CW-OFC beam was set to 1:1 to obtain good contrast for the interferogram in the time domain. A polarization controller (PC) was used for good polarization overlap between CCW-OFC and CW-OFC. The interferogram signal was detected by a fast photodetector (PD1, Thorlabs, Inc., Newton, NJ, USA, PDA015C, bandwidth = DC to 380 MHz) connected to an electric low-pass filter (LPF, cutoff frequency = 21.4 MHz) and was acquired by a digitizer (National Instruments Corp., NI PCI-5122, sampling rate = 100,389,194 samples/s, number of sampling points: 81,353, resolution: 14 bit). A portion of the CW-OFC beam was detected by another fast photodetector (PD2, Thorlabs, Inc., Newton, NJ, USA, PDA015C, bandwidth = DC to 380 MHz) to obtain the RF comb of CW-OFC (freq. = $f_{rep\_CW}$, $2f_{rep\_CW}$, $3f_{rep\_CW}$, •••••, $nf_{rep\_CW}$). Then, we extracted the second harmonic component of $f_{rep\_CW}$ by an electric bandpass filter (BPF, center passband frequency = 87 MHz) and used it as a sampling clock in the digitizer. Although the repetition rate of the interferogram is equal to $\Delta f_{rep}$ (= 1,123 Hz), each interferogram was discretely acquired at 15 Hz in practice to reduce the data size of consecutive interferograms.

## 3. Computational image correction

The residual timing jitter between the dual OFCs in the SCDCFL leads to a blurred image in DCM because the image corresponds to the 2D spectrograph of OFC modes. To compensate for this image blur, we proposed two kinds of image correction based on an image autocorrelation analysis [35]. The first method is self-reference image correction without the use of the reference arm. In this method, we calculated an autocorrelation function of each confocal amplitude image and extracted the center of the function as an error signal of image blur. Then, we corrected the position of each acquired image by comparing the center of the autocorrelation function in the acquired image with that in the first image and compensating for the difference between them.

This method benefits from high robustness to external disturbance due to the use of the common optical path and is effective for imaging static objects with high image contrast. However, it is not suitable for dynamic objects because acquired images are always corrected by comparing images consecutively acquired at different positions with the first image acquired at a fixed position.

To extend the image correction to dynamic objects, we proposed the second method, namely, external-reference image correction using the reference arm. In this case, we simultaneously acquired two confocal amplitude images decoded from the signal-image-encoded CCW-OFC and the reference-image-encoded CCW-OFC. Then, we calculated their autocorrelation functions and extracted their centers as an error signal of image blur. Then, we corrected the position of each acquired signal image by comparing the centers of the autocorrelation functions. While the second image correction is applicable to dynamic samples due to the temporal synchronization between the signal and reference images, the use of a noncommon optical path in the reference arm makes the image correction less robust to external disturbances.

## 4. Results

### 4.1 Basic performance of the bidirectional SCDCFL

We first evaluated the basic performance of the bidirectional SCDCFL. Figure 2(a) shows optical spectra of CCW-OFC and CW-OFC at the oscillator output. The dual OFCs have similar spectral shapes and spectral bandwidths while maintaining good spectral overlap. These spectral characteristics are attractive for DCM. Figures 2(b) and 2(c) show the temporal fluctuation (gate time = 1 s) and the corresponding Allan deviation of $f_{rep\_CCW}$ and $f_{rep\_CW}$, which are indicated by red and blue plots, respectively. In Fig. 2(b), while $f_{rep\_CCW}$ and $f_{rep\_CW}$ have slow drifts, their drifts are common to each other with a constant frequency spacing. Such behavior is also well reflected in Fig. 2(c) as overlapping profiles of the Allan deviation. Figure 2(d) shows the temporal fluctuation of $\Delta f_{rep}$ (gate time = 1 s). Regardless of the slow drifts in $f_{rep\_CCW}$ and $f_{rep\_CW}$, $\Delta f_{rep}$ was passively stabilized within an Allan deviation of 0.0076 Hz at a gate time of 1 s. The Allan deviation of $\Delta f_{rep}$ is shown in Fig. 2(e). Although the laser cavity includes a small noncommon optical path portion, its noncommon-mode influence is negligible.

4.2 Confocal amplitude and phase imaging of static objects

Before performing the computational image correction, we acquired the signal image and the reference image from CCW-OFC, as shown in Figs. 3(a) (image size = 252 µm by 294 µm, pixel size = 50 pixels by 348 pixels, no image accumulation) and 3(b) (image size = 1.00 mm by 1.16 mm, pixel size = 50 pixels by 348 pixels, no image accumulation). Confocal amplitude images of test charts, corresponding to the 2D reflection distribution, were obtained with moderate contrast even in single-shot image acquisition (image acquisition time = 890 µs), although a spatial resolution difference exists between the images due to the different imaging optics. Then, we calculated their image autocorrelation functions, as shown in Figs. 3(c) and 3(d). The left graph of Fig. 3(c) and the right graph of Fig. 3(d) show the amplitude profiles along the vertical white lines in Figs. 3(c) and 3(d), respectively. Using these profiles, we performed self-reference or external-reference image correction.

Figure 4 compares snapshots of confocal amplitude and phase images for a static test chart (image size = 252 µm by 294 µm, pixel size = 50 pixels by 348 pixels, image acquisition time = 890 µs) among (a) no image correction, (b) self-reference image correction, and (c) external-reference image correction. Their corresponding movie is shown in Visualization 1. When comparing these images at the beginning and end of the measurement, the image position moved in the no image correction result of Fig. 4(a) despite the static sample due to the residual timing jitter in the SCDCFL. However, little change was observed in the image position in the self-reference image correction result of Fig. 4(b) and the external-reference image correction result of Fig. 4(c). This effect of the image correction was more clearly confirmed in Visualization 1. In the no image correction, the position of the confocal amplitude and phase images quickly fluctuated along the vertical direction; additionally, its slow drift along the same direction was confirmed (for example, see the moving horizontal black stripe in Visualization 1). However, no fluctuation or drift of the image occurred along the horizontal direction. In 2D-SE, the dispersion directions of the VIPA and the grating are vertical and horizontal in the image, respectively, and the dispersion power of the VIPA is much larger than that of the grating. The resulting 2D spectrograph spatially develops as a vertical zigzag line in the focal plane [5]. Therefore, the image fluctuation and drift in the vertical direction are more sensitive than those in the horizontal direction. Notably, a vertical image shift by a single pixel corresponds to a frequency shift by $f_{rep}$ in the spectral region; hence, DCM is more sensitive to the residual timing jitter than Doppler-broadening or pressure-broadening gas DCS with GHz spectral features. In the self-reference image correction, the image blur was perfectly suppressed, indicating a high image correction capability for static objects. In the external-reference image correction, while the rapid fluctuation of the image position remained, the slow drift was well suppressed. For quantitative analysis of the image correction, we calculated the temporal fluctuation of the center of the image autocorrelation function along the vertical direction from confocal amplitude images of Visualization 1, as shown in Fig. 4(d). The contribution of image correction was clearly confirmed: combined fast fluctuation and slow drift in no image correction, no fast fluctuation and no slow drift in the self-reference image correction, and only the fast fluctuation in the external-reference image correction.

4.3 Evaluation of image accumulation

Image accumulation is often required to improve the signal-to-noise ratio (SNR) or contrast of the image in DCM. However, when the image position fluctuates or drifts due to the residual timing jitter of the SCDCFL, image accumulation will lead to decreased image SNR and/or contrast. To evaluate the effectiveness of the image correction from the viewpoint of image accumulation, we acquired a series of confocal amplitude and phase images of the same sample and accumulated them. Figure 5 compares (i) single image acquisition, (ii) 10-image accumulation, and (iii) 100-image accumulation with (a) no image correction, (b) self-reference image correction, and (c) external-reference image correction. In the no image correction result of Fig. 5(a), the confocal amplitude and phase images became blurred along the vertical direction due to the residual timing jitter, making observation of the chart pattern difficult. While the image accumulation effectively improves the image quality in the self-reference image correction result of Fig. 5(b), slight blurring occurs in the external-reference image correction result of Fig. 5(c).

    For quantitative analysis of the image accumulation contribution in confocal amplitude imaging, we calculated the image SNR from the confocal amplitude images of Figs. 5(a), 5(b), and 5(c). Here, we defined the image SNR as the ratio of the mean to the standard deviation in the bright square region in the top right of each confocal amplitude image. Figure 5(d) shows the image SNR with respect to the number of accumulated images for no image correction, self-reference image correction, and external-reference image correction. Although no accumulation effects were found for the no image correction, the image SNR was improved with increasing number of accumulated images for the self-reference image correction. For the external-reference image correction, although a slight improvement was confirmed for a small number of accumulated images, the image SNR was increased for a large number of accumulated images. These behaviors correspond well to the temporal fluctuations of

the centers of the image autocorrelation functions in Fig. 4(d).

We also calculated the image contrast of the confocal amplitude images when the image contrast was defined as the ratio of the difference to the sum of the maximum and minimum amplitudes across a chart pattern. Since no differences exist in the image contrast along the horizontal direction (not shown), we show here the image contrast along the vertical direction. Figure 5(e) shows a comparison of the image contrast with respect to the number of accumulated images for no image correction, self-reference image correction, and external-reference image correction. We confirmed a similar effect as for the image SNR in Fig. 5(d): a declined contrast for the no image correction and a slight decline for the self-reference image correction and external-reference image correction.

For quantitative analysis of the image accumulation contribution in confocal phase imaging, we calculated the phase stability from the confocal phase images of Figs. 5(a), 5(b), and 5(c). The phase stability was defined as the standard deviation of the temporal phase noise at the image pixels. Figure 5(f) shows the phase stability with respect to the number of accumulated images for no image correction, self-reference image correction, and external-reference image correction. While the phase fluctuation was not suppressed in the no image correction, we confirmed an improved phase stability in the self-reference image correction and external-reference image correction. In this way, we confirmed the effectiveness of image accumulation with the proposed image correction of confocal amplitude and phase images.

### 4.4 Quantitativeness of confocal phase imaging

We next evaluated the quantitativeness of the confocal phase imaging with self-reference image correction. Although the test chart has surface unevenness corresponding to the presence or absence of reflective film, its reflectivity also depends on the presence or absence of the reflective film. To obtain a reflective sample with surface unevenness and constant reflectivity, we formed a thin gold coating on the test chart and used this chart as a confocal phase imaging sample. Figure 6(a) shows the image corresponding to 1000 accumulated confocal phase images. A confocal phase image of the test chart was clearly visualized with high contrast. Figure 6(b) shows a cross-sectional profile of the confocal phase image along the white vertical line in Fig. 6(a). The step profile was determined to be a phase difference of 0.782 rad, corresponding to a step height of 97.1 nm. For comparison, we beforehand determined this step height to be 90 nm by atomic force microscopy (AFM, Hitachi High-Tech, AFM5500M, axial repeatability ≤ 1 nm). In Fig. 5(f), the phase stability of the self-reference image correction was 0.057 rad for 1000 consecutive images, corresponding to an uncertainty of 14.0 nm in the step height. The difference in the step height between DCM and AFM was within the range of this uncertainty. Importantly, axial precision to nanometer order was achieved with the help of the self-reference image correction even though the free-running SCDCFL was used for DCM.

### 4.5 Confocal amplitude and phase imaging of dynamic objects

We extended the computationally image-corrected DCM to a dynamic object. Here, the test chart was laterally and axially moved by a translation stage. Figure 7 compares snapshots of the (i) confocal amplitude and (ii) phase images for a static test chart (image size = 252 μm by 294 μm, pixel size = 50 pixels by 348 pixels, image acquisition time = 890 μs) among (a) no image correction, (b) self-reference image correction, and (c) external-reference image correction. Visualization 2 shows the movie of confocal amplitude and phase images when the test chart was laterally moved. In the no image correction, movement of the test chart was observed with image blur resulting from the residual timing jitter in the SCDCFL. In the self-reference image correction, the image blur was not suppressed because the image correction was performed through comparison with the first image and was hence unsuitable for a moving object. In the external-reference image correction, the image blur was moderately reduced, and the resulting image visualized the movement of the test chart. Although the background of the confocal phase image largely fluctuated in each frame, the relative phase distribution reflects the surface profile of the sample. Visualization 3 shows the movie of confocal amplitude and phase images when the test chart was axially moved. Since the test chart is static in the lateral dimensions, the suppression effect of image blur was more clearly confirmed than in Visualization 2. More importantly, the confocality of the DCM was confirmed in the confocal amplitude images.

## 5. Discussion

We first discuss the reason for the fluctuating center of the image autocorrelation function in the absence of image correction [see the black plot in Fig. 4(d)]. The center fluctuated within a standard deviation of 10 pixels in the short term and a peak-to-peak of 200 pixels in the long term. The former fluctuation corresponds to $10f_{rep\_CCW}$ or 430 MHz in the optical frequency range and $10\Delta f_{rep}$ or 11.23 kHz in the RF region, whereas the latter fluctuation corresponds to $200f_{rep\_CCW}$ or 8.60 GHz in the optical frequency range and $200\Delta f_{rep}$ or 224.6 kHz in the RF region. We consider that such fluctuations result from the residual timing jitter in the SCDCFL as follows: (1) fluctuation of $f_{rep\_CCW}$ ($\delta f_{rep\_CCW}$) in the optical frequency range, (2) fluctuation of $\Delta f_{ceo}$ ($\delta \Delta f_{ceo}$) in the RF range, and (3) fluctuation of $\Delta f_{rep}$ ($\delta \Delta f_{rep}$) in the RF range. From the center optical frequency of 193.4 THz and $f_{rep\_CCW}$ of 43.04 MH, we assume that the number of OFC modes $m$ is 4,493,000. Regarding (1), the $\delta f_{rep\_CCW}$ of 0.002 Hz at a gate time of 1 s [see Fig. 2(c)] is multiplied by $m$ in the optical frequency region; namely, $m\delta f_{rep\_CCW}$ = 8.986 kHz. This contribution to the fast and slow fluctuations in the optical frequency region is negligible. Regarding (2), the $\delta \Delta f_{ceo}$ of 30.5 kHz in the bidirectional SCDCFL [33] contributes to the fluctuation in the RF region. In other words, $\delta \Delta f_{ceo}$ can change the center of the image autocorrelation function within a few tens of pixels, implying a contribution to the slow and/or fast fluctuations. Regarding (3), $m\delta \Delta f_{rep}$ contributes to the fluctuation in the RF region. As $\delta \Delta f_{rep}$ is 0.0076 Hz at a gate time of 1 s [see Fig. 2(e)], $m\delta \Delta f_{rep}$ is estimated to be 34 kHz in the RF region, leading to fluctuations in the center of the image by a few tens of pixels. The contribution of $m\delta\Delta f_{rep}$ is in reasonable agreement with the slow and fast fluctuations in the absence of image correction. In this way, we consider that the slow and fast fluctuations are mainly due to $\delta\Delta f_{ceo}$ and/or $m\delta\Delta f_{rep}$ in the RF region. Although $\delta\Delta f_{ceo}$ shifts the image position while $m\delta\Delta f_{rep}$ fluctuates the image similar to an accordion, observing such a difference in confocal amplitude and phase images is difficult due to the insufficient number of image pixels and/or the large *m* value.

    The external-reference image correction effectively reduces the slow image blur, as shown by the blue plot in Fig. 4(d), revealing its applicability to both static and dynamic objects; however, the fast image blur remains compared with the self-reference image correction. We next discuss the possibility of further reducing the residual fast image blur. To cancel the effect of the residual timing jitter in the SCDCFL, we used a reference image for image correction. Such use of a reference image should suppress the fluctuation caused by $\delta\Delta f_{ceo}$ and/or $m\delta\Delta f_{rep}$; however, it remained. One reason for the residual fluctuation is the noncommon optical path between the signal-image-encoded CCW-OFC and the reference-image-encoded CCW-OFC. Such spatial noncommoness makes image acquisition sensitive to environmental disturbances because air disturbances or vibrations independently influence the optical paths of these two arms. However, the air disturbances or vibrations are considerably slow compared to the data acquisition speed of the interferogram. Actually, a noncommon reference arm worked well in compensating for the phase fluctuation in previous research on DCM [6, 11]. Another reason is the time separation in the interferogram (= 2.67 ns) between the signal-image-encoded CCW-OFC and the reference-image-encoded CCW-OFC. If this time separation is within the range of the mutual coherence length between the dual OFCs, then its influence is negligible. However, the mutual coherence length of the free-running SCDCFL is shorter than that of the fully stabilized dual OFC sources due to the residual timing jitter. We consider that the time separation of 2.67 ns is outside of the mutual coherence length of the SCDCFL. We consider that the fast image blur remaining in the external-reference image correction result is due to the temporal noncommonness rather than to the spatial noncommonness. To achieve temporal commonness, the signal-image-encoded CCW-OFC and the reference-image-encoded CCW-OFC have to be multiplexed with no time separation. Polarization multiplexing is one possible method, although one has to consider the polarization dependence of the 2D-SE.

## 6. Conclusion

We introduced the SCDCFL into DCM to generalize DCM from the viewpoint of reduced complexity of the light source. To compensate for the image blur caused by the residual timing jitter in the SCDCFL, two kinds of computational image correction were applied for confocal amplitude and phase imaging. The self-reference image correction completely suppressed both the slow and fast image blur in the static sample, and its high phase quantitativeness was highlighted by the surface profilometry of a nanometer-order step surface with an uncertainty of 14.0 nm. The external-reference image correction could compensate for the slow image blur and shows high applicability to both static and moving samples. We discussed the potential to further reduce the fast image blur remaining in the external-reference image correction result. This DCM featuring reduced complexity of the light source will expand the application field of DCM in life sciences and industry.

**Funding**

Exploratory Research for Advanced Technology (ERATO), Japan Science and Technology Agency (MINOSHIMA Intelligent Optical Synthesizer Project, JPMJER1304); Japan Society for the Promotion of Science (18H01901, 18K13768, 19H00871); Cabinet Office, Government of Japan (Subsidy for Reg. Univ. and Reg. Ind. Creation); Nakatani Foundation for Advancement of Measuring Technologies in Biomedical Engineering.

**Disclosures**

The authors declare no conflicts of interest.

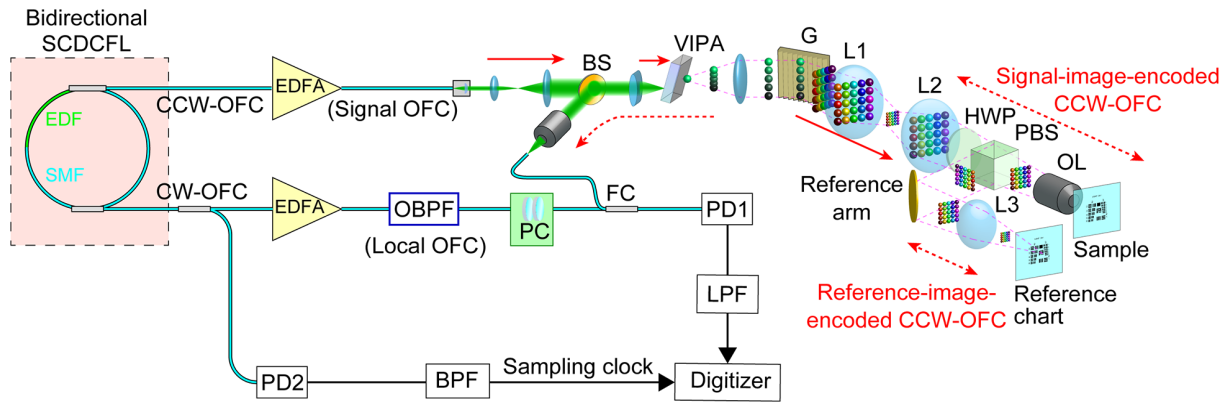

Fig. 1. Experimental setup. EDF, erbium-doped fiber; SMF, single-mode fiber; CCW-OFC, counterclockwise-circulating optical frequency comb; CW-OFC, clockwise-circulating optical frequency comb; EDFAs, erbium-doped fiber amplifiers; BS, beam splitter; VIPA, virtually imaged phased array; G, grating; L1, L2, L3, lenses; HWP, half-wave plate; PBS, polarization beam splitter; OL, objective lens; OBPF, optical bandpass filter; PC, polarization controller; FC, fiber coupler; PD1, PD2, fast photodetectors; LPF, low-pass filter; BPF, bandpass filter.

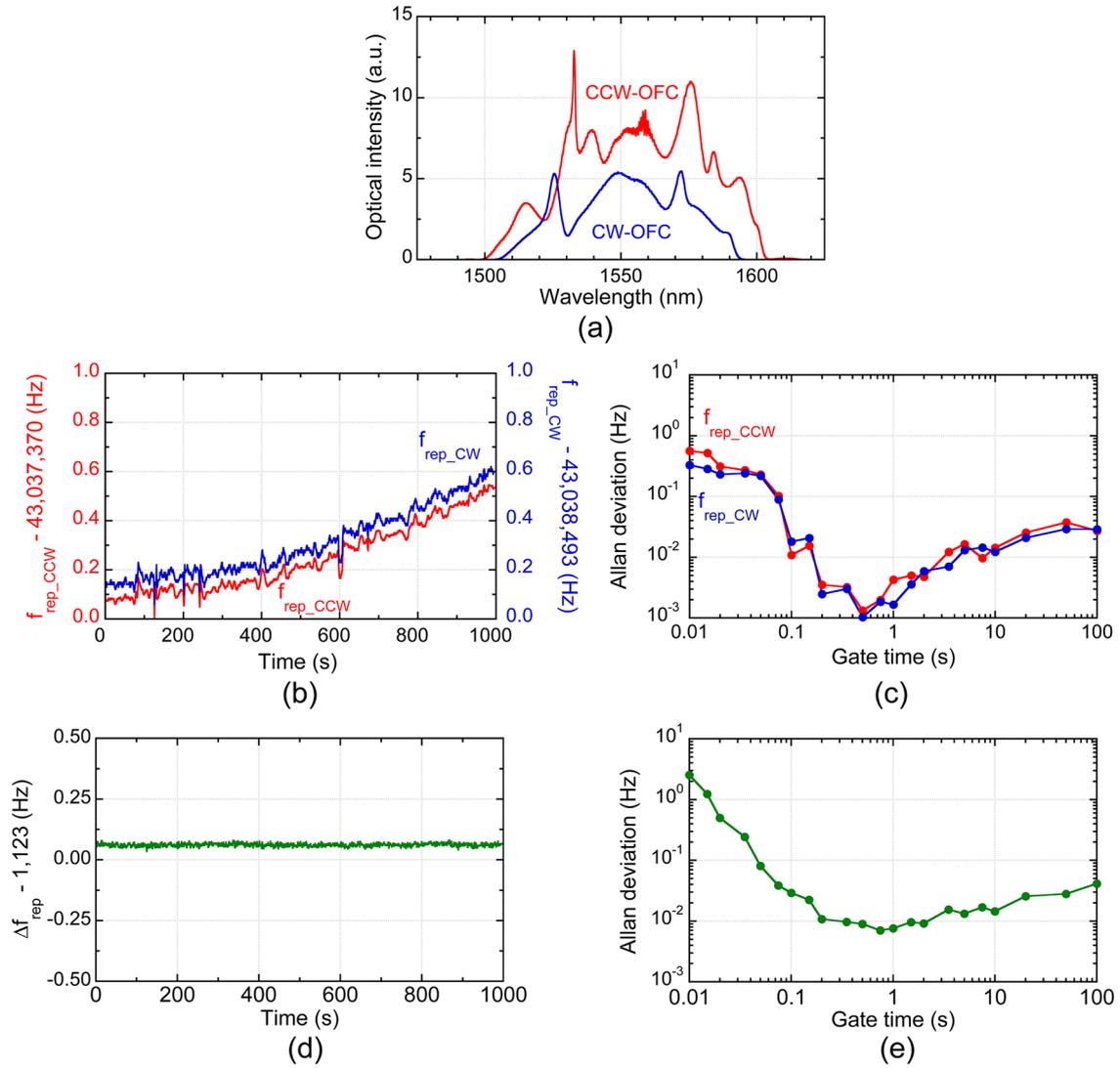

Fig. 2. (a) Optical spectra of CCW-OFC and CW-OFC at the oscillator output. (b) Temporal fluctuation and (c) Allan deviation of $f_{rep\_CCW}$ and $f_{rep\_CW}$. (d) Temporal fluctuation and (e) Allan deviation of $\Delta f_{rep}$.

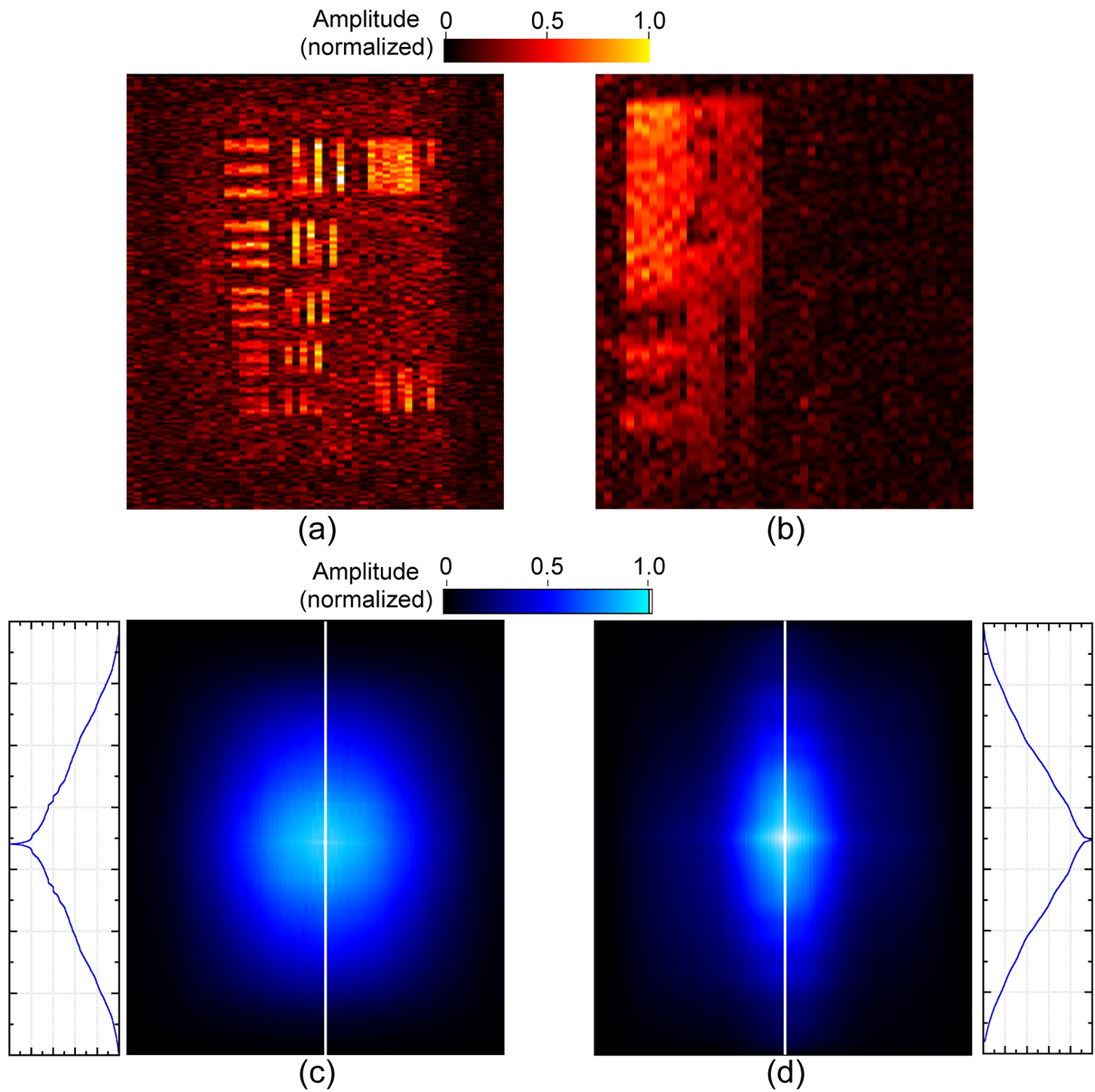

Fig. 3. Confocal amplitude images of a (a) signal test chart and (b) reference test chart. Image autocorrelation functions of the (c) signal test chart and (d) reference test chart along with the amplitude profiles along the vertical white line.

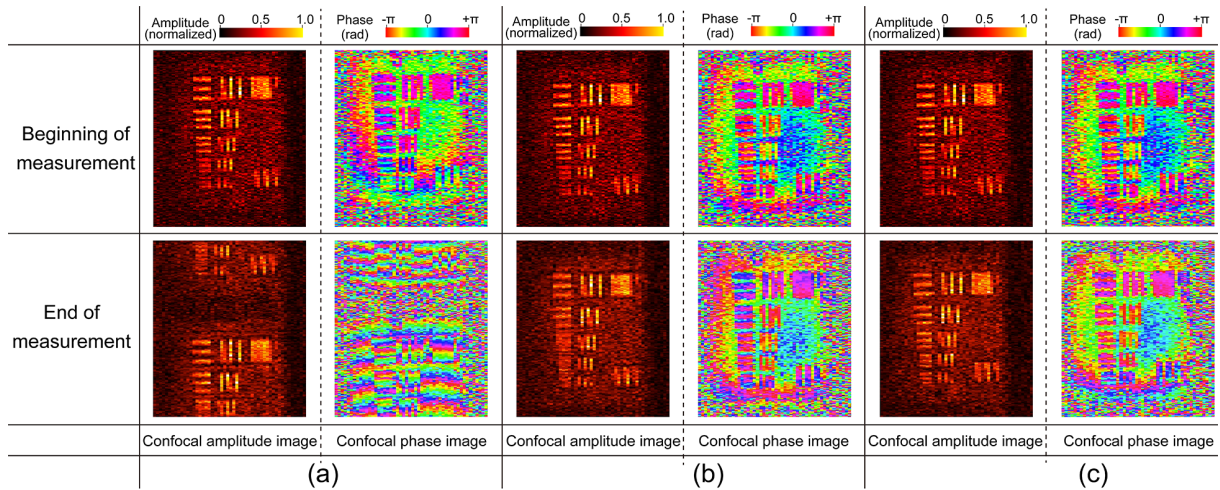

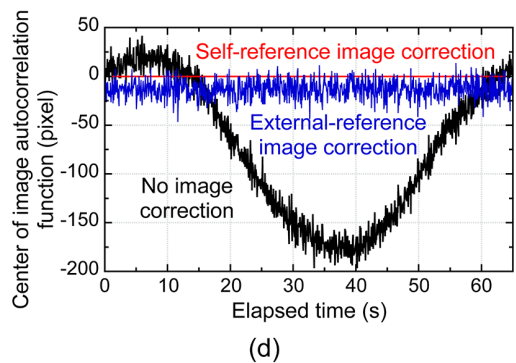

Fig. 4. Snapshots of confocal amplitude and phase images of a static test chart with (a) no image correction, (b) self-reference image correction, and (c) external-reference image correction (see Visualization 1). (d) Comparison of the temporally fluctuating centers of the image autocorrelation functions among no image correction, self-reference image correction, and external-reference image correction.

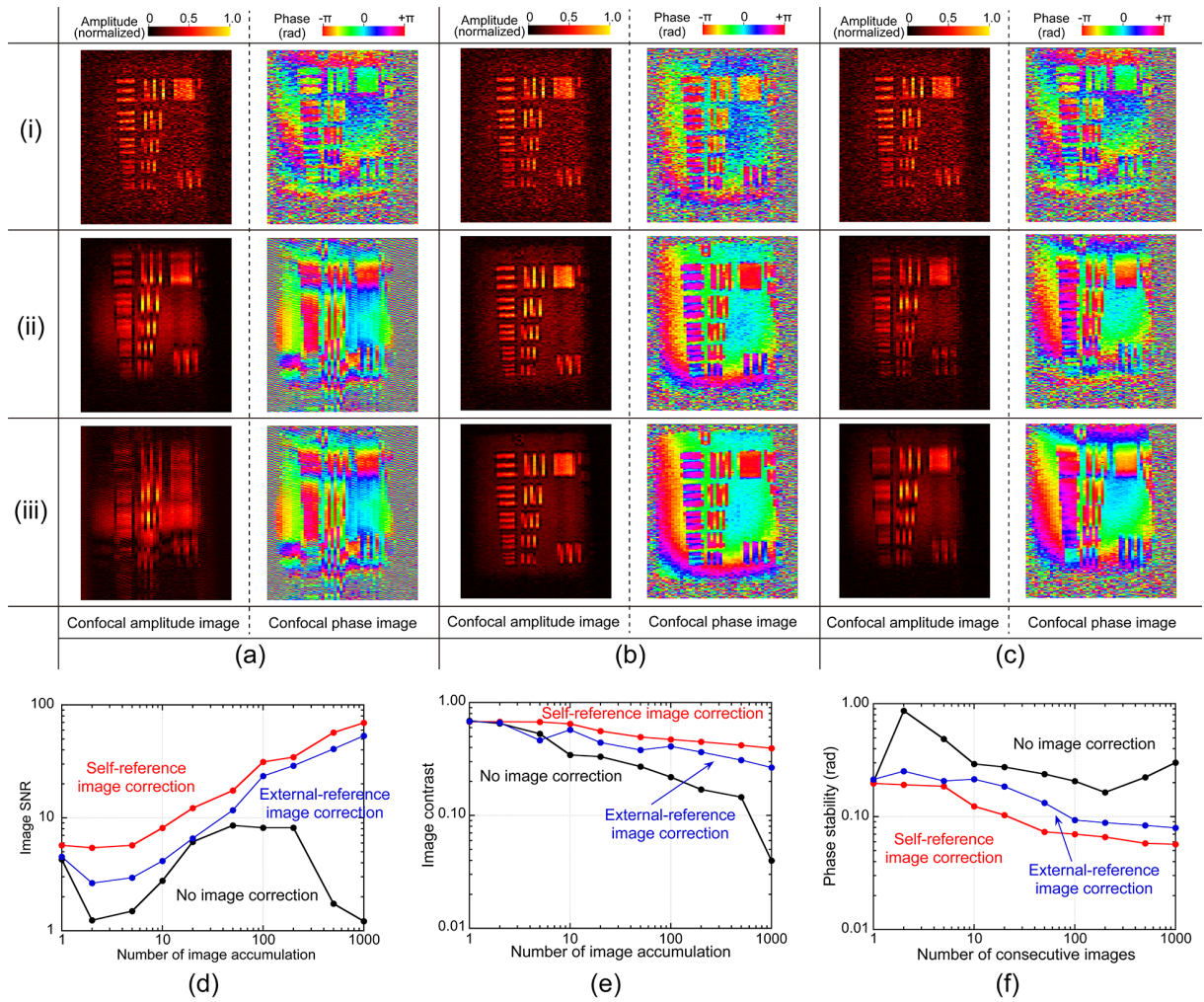

Fig. 5. Comparison of confocal amplitude and phase images: (i) single image acquisition, (ii) 10-image accumulation, and (iii) 100-image accumulation with (a) no image correction, (b) self-reference image correction, and (c) external-reference image correction. Comparison of the (d) image SNR, (e) image contrast, and (f) phase stability with respect to the number of accumulated images among no image correction, self-reference image correction, and external-reference image correction.

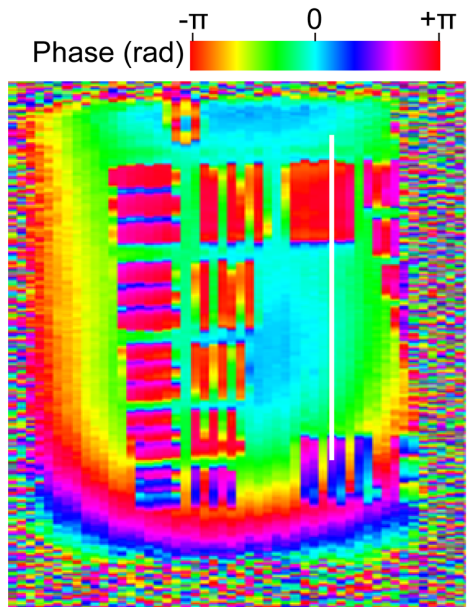 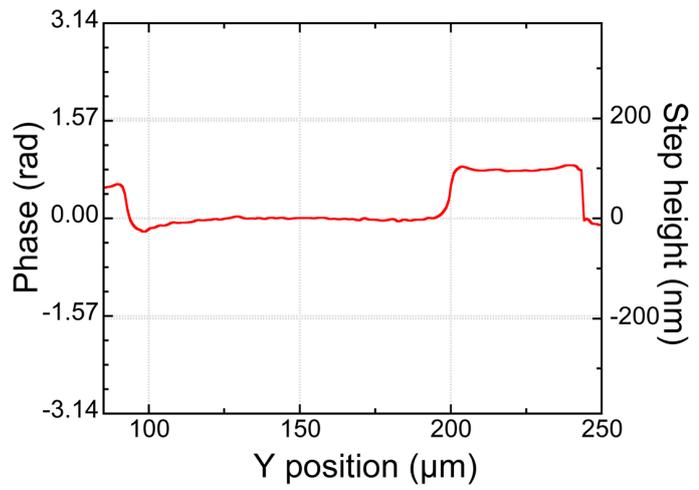

Fig. 6. (a) Image corresponding to 1000 accumulated confocal phase images of a static test chart, and (b) cross-sectional profile of the confocal phase image along the white vertical line in Fig. 6(a).

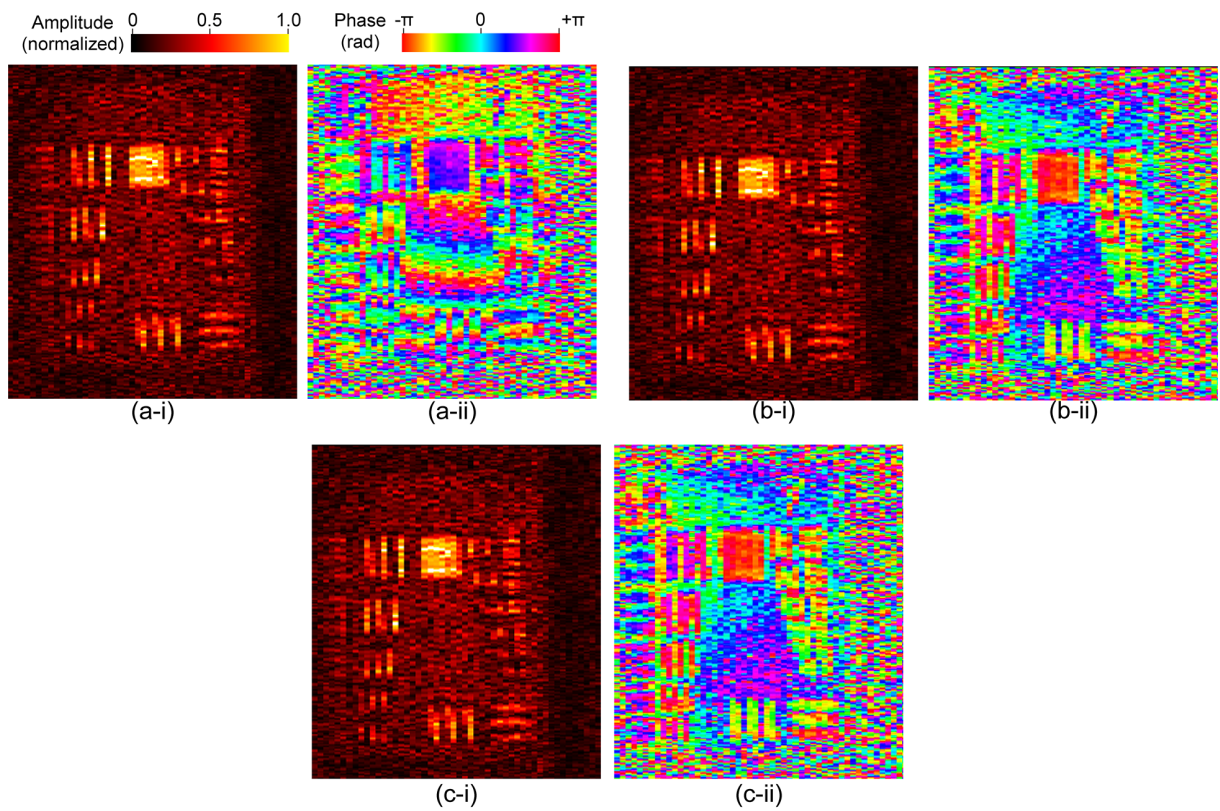

Fig. 7. Comparison of confocal amplitude and phase images: (i) confocal amplitude image and (ii) phase images for a static test chart with (a) no image correction, (b) self-reference image correction, and (c) external-reference image correction (see Visualization 2, Visualization 3).